\documentclass[%
reprint,
superscriptaddress,
amsmath,amssymb,
aps,
prb,
]{revtex4-2}
\usepackage{graphicx,afterpage}
\usepackage{xcolor}
\usepackage{soul}
\usepackage{ulem}
\usepackage[T1]{fontenc} 
\usepackage[colorlinks=true,plainpages=false,linkcolor=blue,urlcolor=blue,citecolor=blue,pdfpagemode=UseNone,pdfstartview=FitBH]{hyperref}
\usepackage{amsmath}
\usepackage[french]{babel}
\usepackage{siunitx}  
\DeclareSIUnit\angstrom{\text {Å}}
\begin{document}


\title{Interplay of Charge and Magnetic Orders in SmNiC$_2$ Mediated by Electron--Phonon Interaction}

\author{A. von Ungern-Sternberg Schwark}
\thanks{Present address: Max Planck Institute for Solid State Research, Heisenbergstrasse 1, 70569 Stuttgart, Germany}
\affiliation{Institute for Quantum Materials and Technologies, Karlsruhe Institute of Technology, Kaiserstraße 12, 76131 Karlsruhe, Germany}
\author{A.-A. Haghighirad}
\affiliation{Institute for Quantum Materials and Technologies, Karlsruhe Institute of Technology, Kaiserstraße 12, 76131 Karlsruhe, Germany}
\author{R. Heid}
\affiliation{Institute for Quantum Materials and Technologies, Karlsruhe Institute of Technology, Kaiserstraße 12, 76131 Karlsruhe, Germany}
\author{P. H. McGuinness}
\affiliation{Institute for Quantum Materials and Technologies, Karlsruhe Institute of Technology, Kaiserstraße 12, 76131 Karlsruhe, Germany}
\author{N. Maraytta}
\affiliation{Institute for Quantum Materials and Technologies, Karlsruhe Institute of Technology, Kaiserstraße 12, 76131 Karlsruhe, Germany}
\author{A. Eich}
\affiliation{Institute for Quantum Materials and Technologies, Karlsruhe Institute of Technology, Kaiserstraße 12, 76131 Karlsruhe, Germany}
\author{M. Merz}
\affiliation{Institute for Quantum Materials and Technologies, Karlsruhe Institute of Technology, Kaiserstraße 12, 76131 Karlsruhe, Germany}
\affiliation{Karlsruhe Nano Micro Facility (KNMFi), Karlsruhe Institute of Technology, Kaiserstraße 12, 76131 Karlsruhe, Germany\\}

\author{A. Bosak}
\affiliation{ESRF – The European Synchrotron, 71, avenue des Martyrs, CS 40220 F-38043 Grenoble Cedex 9}
\author{D. A. Chaney}
\affiliation{ESRF – The European Synchrotron, 71, avenue des Martyrs, CS 40220 F-38043 Grenoble Cedex 9}
\author{A. Chumakova}
\affiliation{ESRF – The European Synchrotron, 71, avenue des Martyrs, CS 40220 F-38043 Grenoble Cedex 9}

\author{A. Pawbake}
\affiliation{Laboratoire National des Champs Magn{\'e}tiques Intenses, CNRS-UGA-UPS-INSA-EMFL, 38042 Grenoble, France}
\author{C. Faugeras}
\affiliation{Laboratoire National des Champs Magn{\'e}tiques Intenses, CNRS-UGA-UPS-INSA-EMFL, 38042 Grenoble, France}

\author{M. Le Tacon}
\email{matthieu.letacon@kit.edu}	
\affiliation{Institute for Quantum Materials and Technologies, Karlsruhe Institute of Technology, Kaiserstraße 12, 76131 Karlsruhe, Germany}
\author{S. M. Souliou}
\email{michaela.souliou@kit.edu}
\affiliation{Institute for Quantum Materials and Technologies, Karlsruhe Institute of Technology, Kaiserstraße 12, 76131 Karlsruhe, Germany}

\begin{abstract}

We investigate the interplay between charge density wave (CDW) instabilities and ferromagnetism in SmNiC$_2$ using diffuse and inelastic x-ray scattering together with Raman spectroscopy. We identify a soft acoustic phonon driving the incommensurate CDW (I-CDW) and uncover a second Kohn anomaly at the wave vector of the commensurate CDW (C-CDW) stabilized in other $R$NiC$_2$ members ($R=$\,rare earth). The marked softening of both phonons and their contrasting evolution with temperature reveal a competition between the two ordering tendencies. Alongside pronounced anomalies in the temperature dependence of the zone center and soft phonons, we observe the collective amplitude mode of the CDW, which collapses abruptly as ferromagnetism sets in. Surprisingly, the Kohn anomalies persist in the ferromagnetic state despite the degradation of Fermi-surface nesting conditions. Our experimental findings, supported by \textit{ab initio} calculations, highlight the central role of the electron--phonon interaction in driving the CDW formation and tuning the balance between competing charge and magnetic orders.

\end{abstract}

\maketitle

\textit{Introduction --} Charge density wave (CDW) phases arise in a wide range of quantum materials, often coexisting, competing or intertwined with other electronic orders. Their interplay with superconductivity has been intensively studied in transition metal dichalcogenides~\cite{Chen2016}, cuprates~\cite{Frano2020}, and kagome metals~\cite{Wilson2024}. Less commonly, CDWs occur alongside magnetic order, either as modulations within a magnetically ordered state, as in FeGe~\cite{Teng2023,Wu2024}, or as a host lattice for magnetic order, as in EuAl$_{4}$~\cite{Shimomura2019}. Understanding how lattice, charge, and spin degrees of freedom interrelate in such systems and in particular clarifying the role of electron--phonon coupling (EPC), traditionally linked to CDW formation~\cite{Johannes2008}, remain central challenges.

A fertile setting for addressing this issue is offered by the non-centrosymmetric rare-earth nickel dicarbides, $R$NiC$_2$, which host multiple CDW, magnetic, and superconducting phases~\cite{Roman2018a,Maeda2019,Landaeta2017,Katano2019}. Mainly two distinct CDWs are observed: an incommensurate order (I-CDW) with wavevector ${\boldsymbol{q}_\mathrm{I}}$ = (0.5, 0.5+$\xi$, 0), characteristic of larger-unit-cell members such as NdNiC$_2$ and SmNiC$_2$~\cite{Yamamoto2013,Shimomura2009}, and a commensurate order (C-CDW) with ${\boldsymbol{q}_\mathrm{C}}$ = (0.5, 0.5, 0.5), found in smaller-cell members such as ErNiC$_2$ and TmNiC$_2$~\cite{Maeda2019,Roman2023}. Some intermediate compounds display both instabilities and/or lock-in behavior, e.g. TbNiC$_2$~\cite{Shimomura2016}. At low temperature, the rare-earth moments in most $R$NiC$_2$ compounds order antiferromagnetically, with the exception of SmNiC$_2$, which undergoes a first-order ferromagnetic transition at $T_\mathrm{C}=17.5$\,K~\cite{Onodera1998}.

\begin{figure*}[!ht]
\includegraphics[width=1\textwidth]{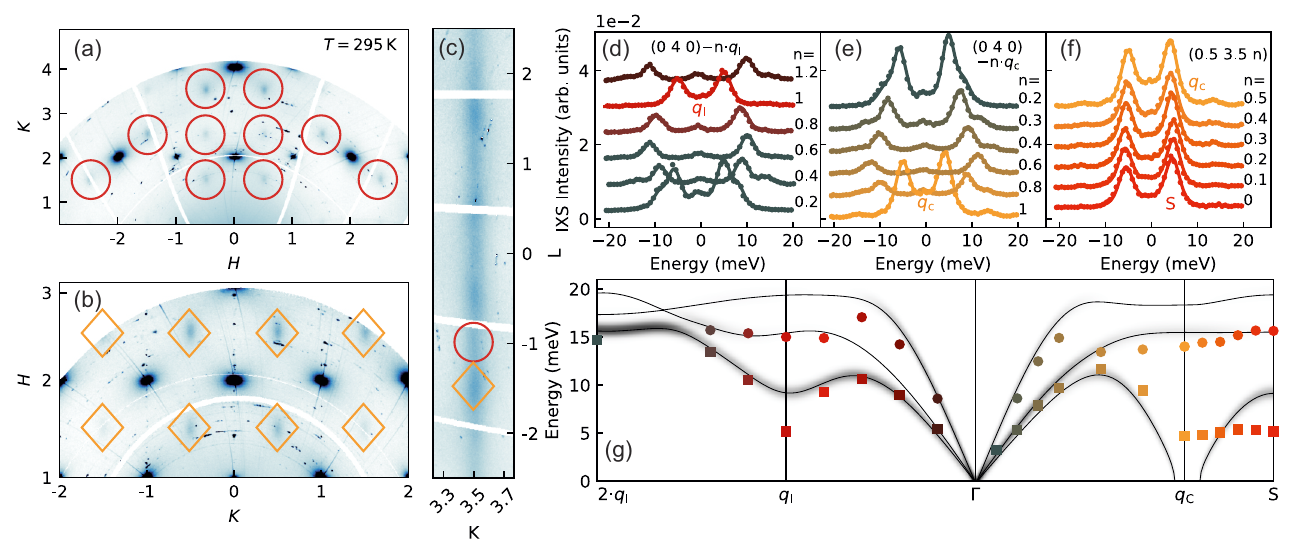}
\centering
\caption{(a--c) Room temperature reciprocal space maps of the (a) ($HK$0), (b) ($HKK$), and (c) (0.5$KL$) planes of SmNiC$_2$. DS signal at $\boldsymbol{q}_\mathrm{I}$ ($\boldsymbol{q}_\mathrm{C}$) is indicated by red (mustard) circles (rhombi). (d--f) Room temperature IXS spectra taken in the reciprocal space paths from $\bf{\Gamma_{040}}$ (d) along ${\boldsymbol{q}_\mathrm{I}}$ and (e) along $\boldsymbol{q}_\mathrm{C}$ and (f) from $\boldsymbol{q}_\mathrm{C}$ along the [00$l$] direction. The S point is very close to $\boldsymbol{q}_\mathrm{I}$ (neglecting the small incommensurability). The lines correspond to data fits. A vertical offset is included. (g) Phonon dispersion along the paths of the spectra shown in (d--f), with a shared color scheme. The points correspond to the experimentally determined dispersion at room temperature and the lines are the results of DFPT calculations for LaNiC$_2$, including a grayscale representation of the calculated structure factors~\cite{SOM}.}
\label{Fig1}
\end{figure*}

SmNiC$_2$ is particularly intriguing. X-ray diffraction (XRD) established an I-CDW below $T_\mathrm{CDW}=148$\,K~\cite{Shimomura2009}, accompanied by diffuse scattering (DS) signal near $\boldsymbol{q}_\mathrm{C}$ but without long-range order. Structural refinements identified Ni displacements along the $a$-axis as the main CDW distortion~\cite{Woelfel2010}. Remarkably, both the I-CDW satellites and the diffuse signal vanish abruptly below $T_\mathrm{C}$, concomitant with a sharp resistivity drop~\cite{Shimomura2009}. Magnetic field tuning further confirmed the mutual exclusion of CDW and ferromagnetic order~\cite{Hanasaki2012}, in contrast to NdNiC$_2$ and GdNiC$_2$, where antiferromagnetism coexists with the CDW~\cite{Yamamoto2013,Shimomura2016}.

Electronic structure calculations support a nesting-driven CDW formation mechanism: pronounced susceptibility peaks appear near $\boldsymbol{q}_\mathrm{I}$ and more weakly near $\boldsymbol{q}_\mathrm{C}$, which deteriorate upon entering the ferromagnetic phase due to Fermi surface spin splitting~\cite{Laverock2009,Kim2013}. DS around both vectors has been interpreted as evidence for Kohn anomalies~\cite{Shimomura2009,Shimomura2016}, yet direct phonon measurements are lacking. Experimental phonon studies across the Brillouin zone or at the zone center have not been reported thus far for SmNiC$_2$ or other $R$NiC$_2$ members. Consequently, confirmation of the phonon response to the underlying CDW and magnetic competition has remained elusive.

Here we address this issue using diffuse and inelastic x-ray scattering (IXS) together with Raman spectroscopy to probe low-energy excitations in SmNiC$_2$. We directly identify the soft-phonon origin of the I-CDW, reveal a temperature-driven shift of the dominant phonon anomaly from ${\boldsymbol{q}_\mathrm{C}}$ to ${\boldsymbol{q}_\mathrm{I}}$, and establish the competition between the two CDW tendencies. Upon entering the ferromagnetic phase, the dips in the dispersion persist even as the I-CDW order is suppressed, an effect corroborated by the disappearance of the CDW amplitude mode and back-folded phonons in Raman spectra. Our findings highlight the role of EPC in mediating the unusual charge and magnetic interplay in SmNiC$_2$.

\textit{Methods --} Single crystals of SmNiC$_2$ were grown using argon arc-melting. In agreement with earlier reports, our structural analysis using XRD confirmed the non-centrosymmetric orthorhombic CeNiC$_2$-type structure ($Amm$2). DS and IXS experiments were performed at the ID28 beamline of the ESRF~\cite{Krisch2006,Girard2019}. Raman scattering experiments were performed with a LabRAM HR Evolution spectrometer equipped with a nitrogen-cooled CCD camera. Further information on the crystal growth, structural characterization, experimental and computational methods can be found in the Supplemental Material~\cite{SOM}.

\textit{X-ray scattering --} In addition to the Bragg reflections described in the space group $Amm$2, and in line with earlier studies~\cite{Shimomura2009,Shimomura2016}, DS signal is observed around ${\boldsymbol{q}_\mathrm{I}}$ and ${\boldsymbol{q}_\mathrm{C}}$ already at ambient conditions [Fig.~\ref{Fig1}(a--c)]. In the ($HK$0) and ($HKK$) planes this appears as periodic, diffuse, oval-shaped features in between Bragg nodes. In the (0.5$KL$) plane -- in which there are no Bragg peaks of the fundamental structure -- the diffuse signal appears as rods running along the $l$ direction, i.e. along the direction connecting ${\boldsymbol{q}_\mathrm{I}}$ and ${\boldsymbol{q}_\mathrm{C}}$ (neglecting the small incommensurability of $\boldsymbol{q}_\mathrm{I}$). Interestingly, at ambient conditions the intensity maxima of these rods are centered around $\boldsymbol{q}_\mathrm{C}$, i.e. at $l=0.5$ and not around $\boldsymbol{q}_\mathrm{I}$ at $l=0$. 

\begin{figure}
\centering
\includegraphics[width=0.45\textwidth]{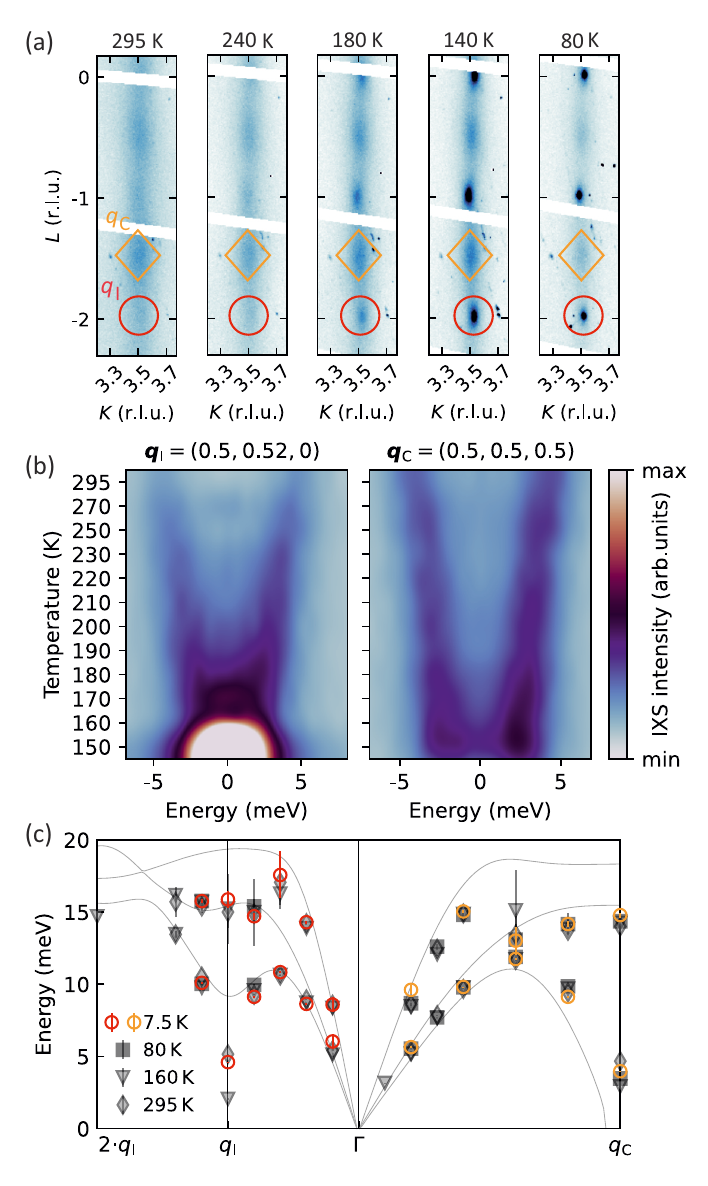}
\caption{(a)  Reciprocal space maps of the (0.5$KL$) plane of SmNiC$_2$ at low temperatures. (b) Colormap of the temperature dependence of the IXS spectra recorded at $\boldsymbol{q}_\mathrm{I}$ and $\boldsymbol{q}_\mathrm{C}$ upon approaching $T_ \mathrm{CDW}$. (c) Temperature dependence of the phonon dispersion along $\boldsymbol{q}_\mathrm{I}$ and $\boldsymbol{q}_\mathrm{C}$ (red and orange symbols respectively at 7.5 K). The points correspond to experimental results and the lines to DFPT calculations~\cite{SOM}.}
\label{Fig2}
\end{figure}	

To test the origin of the DS signal, we performed energy-resolved IXS experiments close to the $\bf{\Gamma_{040}}$ Bragg. The spectra [Fig.~\ref{Fig1}(d--f)] consist of a weak quasi-elastic line and two phonon modes below 20\,meV. Along $\boldsymbol{q}_\mathrm{I}$ and $\boldsymbol{q}_\mathrm{C}$, the lowest-energy acoustic branch dominates the scattering intensity and softens anomalously around both ordering vectors. Fits using damped harmonic oscillator functions yield the experimental dispersion [Fig.~\ref{Fig1}(g)]. The soft branch shows minima at $\boldsymbol{q}_\mathrm{I}$ ($\sim$5.2\,meV) and $\boldsymbol{q}_\mathrm{C}$ ($\sim$4.7\,meV), and remains nearly flat between them, consistent with the diffuse rods shown in Fig.~\ref{Fig1}(c). These results establish that the DS signal originates from low-energy phonons~\cite{Bosak2015}.
Density-functional perturbation theory (DFPT) calculations of the phonon dispersion and structure factors for LaNiC$_2$~\cite{SOM} match the experiment well, but predict the dominant instability at $\boldsymbol{q}_\mathrm{C}$.

The temperature evolution of the DS signal and IXS spectra at $\boldsymbol{q}_\mathrm{I}$ and $\boldsymbol{q}_\mathrm{C}$ is shown in Fig.~\ref{Fig2}. Upon cooling, the DS signal sharpens and intensifies around both vectors. The intensity grows stronger at $\boldsymbol{q}_\mathrm{I}$ and condenses into satellites of the I-CDW state below $T_\mathrm{CDW}$. The signal at $\boldsymbol{q}_\mathrm{C}$ remains diffuse and is partially suppressed in the I-CDW phase~\cite{Shimomura2009}. IXS reveals that the lowest-energy phonons soften upon cooling at both $\boldsymbol{q}_\mathrm{I}$ and $\boldsymbol{q}_\mathrm{C}$, with the softening however being stronger at $\boldsymbol{q}_\mathrm{I}$. At this wavevector, the spectra below 170\,K are reduced to a quasielastic peak that intensifies and narrows upon approaching $T_\mathrm{CDW}$, evidencing the complete phonon softening. At $\boldsymbol{q}_\mathrm{C}$, the phonons remain resolvable, softening to $\sim$2.9\,meV at $T_\mathrm{CDW}$, before rehardening to $\sim$3.5\,meV at 110\,K and displaying a weak temperature dependence upon further cooling towards $T_\mathrm{C}$~\cite{SOM}. The inverse relation between DS intensity and phonon energy further supports the phononic origin of the DS signal.

In the ferromagnetic state, the phonon energy at $\boldsymbol{q}_\mathrm{C}$ abruptly increases (from \textit{E}$\sim$3.6\,meV at 22\,K to \textit{E}$\sim$4\,meV at 12\,K -- Fig.~\ref{Fig4}~\cite{SOM}), while the destruction of the I-CDW order~\cite{Shimomura2009} allows investigations of the low energy phonons also at $\boldsymbol{q}_\mathrm{I}$, i.e. at the zone center of the suppressed I-CDW. At this wavevector the phonon re-emerges with \textit{E}$\sim$4.7\,meV at 12\,K. Importantly, the phonon dips around both $\boldsymbol{q}_\mathrm{I}$ and $\boldsymbol{q}_\mathrm{C}$ persist below $T_\mathrm{C}$, despite the suppression of long-range CDW order (Fig.~\ref{Fig2}(c))~\cite{SOM}.
 
\begin{figure}[t]
\centering
\includegraphics[width=0.45\textwidth]{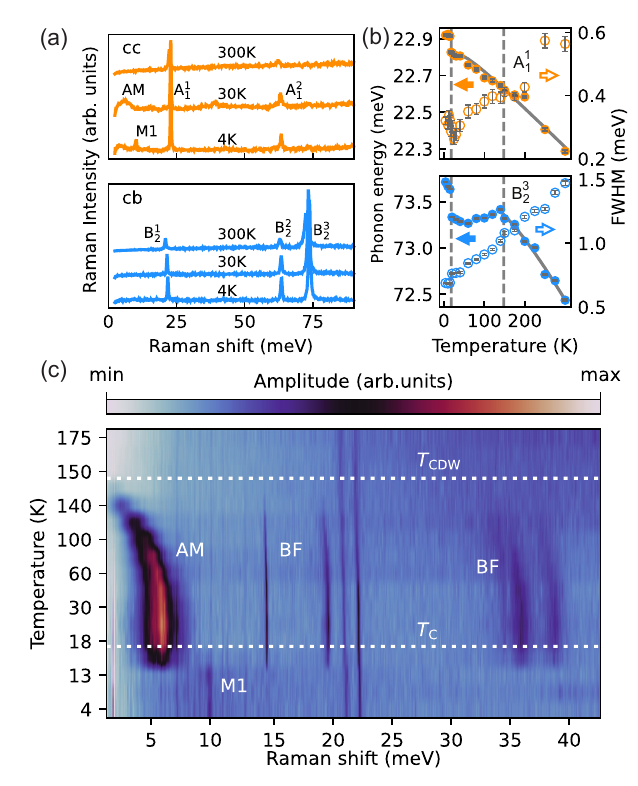}
\caption{(a) Raman spectra of SmNiC$_2$ at 300, 30 and 4\,K recorded in the $cc$ and $cb$ scattering geometries. The spectra are vertically shifted. Phonons of $A_1$ and $B_2$ symmetry are indicated. (b) Temperature dependence of the energy (closed circles) and full width at half maximum (FWHM, open circles) of the $A_1^1$ and $B_2^3$ phonons labeled in (a). (c) Colormap of the temperature dependence of the spectra recorded in the $bb$ scattering geometry. AM and BF refer to the CDW amplitude mode and back-folded phonons respectively, whereas M1 denotes the feature observed in the ferromagnetic phase.} 
\label{Fig3}
\end{figure}	

\textit{Raman scattering --} To complement the phonon dispersion studies, we performed zone-center Raman measurements across the CDW and ferromagnetic transitions (Fig.~\ref{Fig3}). SmNiC$_2$ has nine optical modes ($\Gamma$ = 3$A_1$ + $A_2$ + 2$B_1$ + 3$B_2$). In our experiments, the incident and scattered light polarizations were within the $bc$ plane, allowing observation of all $A_1$ and $B_2$ phonons with frequencies consistent with calculations~\cite{SOM}. 

Pronounced new Raman features emerge in the I-CDW phase: a low-energy mode ($\sim$5.5\,meV at 30\,K) that hardens, sharpens and gains intensity upon cooling below $T_\mathrm{CDW}$ and higher-energy modes ($\sim$15, 20 and 35--40\,meV) that remain nearly temperature-independent. These features all have $A_1$ symmetry and are consistent with a CDW amplitude mode (AM) and back-folded phonon modes respectively.

Crossing into the ferromagnetic phase, all CDW-related features vanish abruptly, while most of the intrinsic phonons undergo a sudden blueshift, evidencing magneto-structural coupling in agreement with our low temperature XRD measurements and earlier reports revealing anomalies in the lattice parameters across the first-order transition~\cite{Murase2004,SOM}. A new mode emerging near 10\,meV in the magnetic phase, which renormalizes under magnetic field~\cite{SOM}, points to a distinct magnetic excitation that will be addressed in future work.

\begin{figure}[t]
\centering
\includegraphics[width=0.45\textwidth]{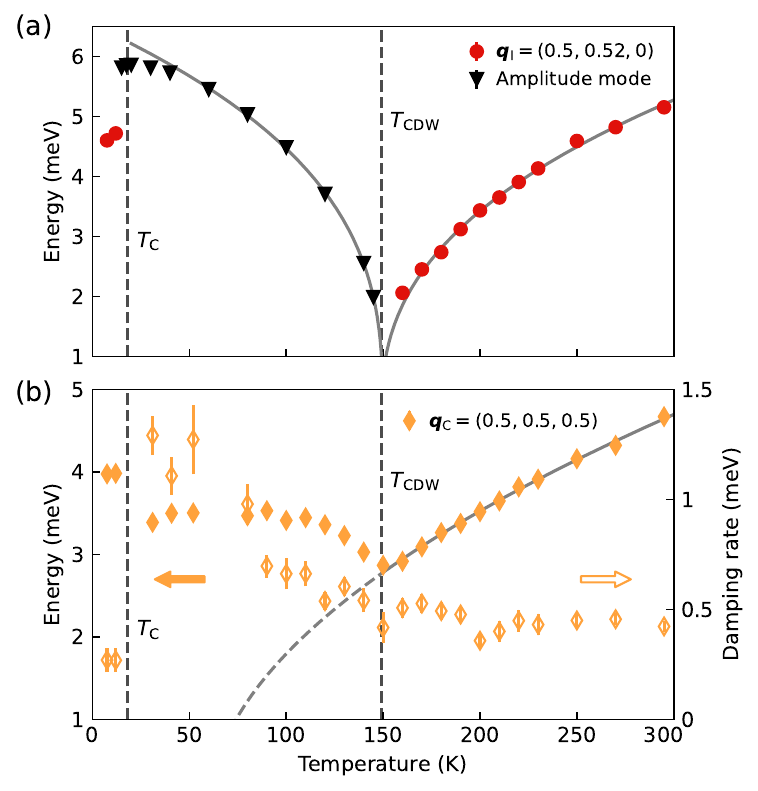}
\caption{(a) Temperature dependence of the soft phonon and amplitude mode energies in Sm$\mathrm{Ni}\mathrm{C}_{2}$. (b) Energy and damping rate of the phonon at $\boldsymbol{q}_\mathrm{C}$. The soft phonons at $\boldsymbol{q}_\mathrm{I}$ and $\boldsymbol{q}_\mathrm{C}$ were measured with IXS for $T>T_\mathrm{CDW}$ and $T<T_\mathrm{C}$, whereas the amplitude mode was measured with Raman scattering for $T_\mathrm{C}<T<T_\mathrm{CDW}$.} 
\label{Fig4}
\end{figure}	

\textit{Discussion --}
We begin our discussion addressing the CDW formation mechanism in SmNiC$_2$. Our experimental results unambiguously establish that the transition to the I-CDW phase is driven by the softening of an acoustic phonon. The relevant phonon branch is soft along the entire path connecting $\boldsymbol{q}_\mathrm{I}$ and $\boldsymbol{q}_\mathrm{C}$, with minima around these two wave vectors, manifesting the tendency towards both types of ordering. Moreover, the detailed temperature dependence reveals a shift in the wave vector of the global phonon minimum. At room temperature the branch is softest at $\boldsymbol{q}_\mathrm{C}$ -- indicating C-CDW as the leading ordering tendency -- however, the phonon at $\boldsymbol{q}_\mathrm{I}$ softens more rapidly upon cooling, rendering the instability towards I-CDW as the dominant one below $\sim$210\,K and freezing into the lattice at $T_\mathrm{CDW}$. Interestingly, the formation of the I-CDW order interrupts the continuous phonon softening at $\boldsymbol{q}_\mathrm{C}$, evidencing the hindered tendency towards C-CDW and the competition between I-CDW and C-CDW ordering.

Our experimental observations are consistent with calculations of the electronic susceptibility and the EPC both displaying maxima around $\boldsymbol{q}_\mathrm{I}$ and $\boldsymbol{q}_\mathrm{C}$~\cite{Laverock2009,Kim2013,Roman2025}. The prevailing I-CDW order is the result of the combined effect of both contributions, alongside effects of lattice anharmonicity. Naturally, the stabilization of I-CDW modifies the electronic structure and -- necessarily -- also the EPC, affecting the tendency towards C-CDW ordering. Indeed, recent susceptibility calculations for YNiC$_2$ in the I-CDW modulated structure show a broad and shallow feature at $\boldsymbol{q}_\mathrm{C}$, consistent with the phonon hardening across $T_\mathrm{CDW}$ observed here~\cite{Roman2025}. The fact that the Kohn anomaly at $\boldsymbol{q}_\mathrm{C}$ is only partially weakened is moreover in line with calculations of locally enhanced EPC at this wave vector in the I-CDW phase~\cite{Roman2025}. We note that in other ${R}$NiC$_2$ members, e.g. DyNiC$_2$, the C-CDW order is stabilized at temperatures lower than the I-CDW onset~\cite{Maeda2019}, indicating a complete phonon softening at $\boldsymbol{q}_\mathrm{C}$, and therefore a modified balance of the two CDW instabilities.

The detailed temperature dependence of the soft phonons and the collective amplitude mode of the CDW is summarized in Fig.~\ref{Fig4}. Unlike the back-folded phonon modes which show a modest temperature dependence, the AM energy can be described by a modified mean field expression
\begin{equation}\label{eq1}
E_\mathrm{AM}=E_0(1-T/T_\mathrm{CDW})^{\gamma}
\end{equation}
with $E_0=6.5\pm0.05\,$meV, $T_\mathrm{CDW}=149.8\pm 0.5\,$K and $\gamma=0.35\pm0.01$ (Fig.~\ref{Fig4}). The resulting $\gamma$ is close to values for the universal critical exponent of the order parameter found in second-order phase transitions in 3D systems~\cite{Chaikin_Lubensky_book}. The energy of the AM surpasses that of the soft mode at $\boldsymbol{q}_\mathrm{C}$ at $\sim$130\,K and saturates at $\sim$5.83\,meV just above $T_\mathrm{C}$. Interestingly, the soft phonon at $\boldsymbol{q}_\mathrm{C}$ significantly broadens below $T_\mathrm{CDW}$ (along with hardening, Fig.~\ref{Fig4})~\cite{SOM}, corresponding to higher damping rate and shorter phonon lifetime in the I-CDW state. This could relate to increased electron--phonon decay rates due to enhanced local EPC~\cite{Roman2025}. Alternatively, it could arise from additional phonon decay pathways, such as coupling to collective I-CDW excitations rooted in EPC.

Pronounced anomalies across $T_\mathrm{CDW}$ are also observed for the zone center $A_1$ and $B_2$ phonons of the main structure. As shown in Fig.~\ref{Fig3} and detailed in~\cite{SOM}, their energy and linewidths display clear renormalizations across $T_\mathrm{CDW}$. Interestingly, this is most prominent for phonons with energies close to that of the (partial) gap 
at the Fermi level revealed by photoemission ($\sim$60--70\,meV~\cite{Sato2010}). These phonons also have asymmetric Fano and/or broad ($\sim$1--1.5\,meV at ambient conditions) lineshapes, both signatures of coupling to the underlying electronic continuum. 
Their calculated phonon linewidths arising from the EPC are larger than those of lower energy modes~\cite{SOM}. Structural modifications in the modulated phase could result in the softening/broadening of phonons whose vibrational patterns are most affected. However, the energy scale of the most pronounced phonon effects instead suggests a phononic response to the electronic gap opening, akin to renormalizations seen across superconducting transitions~\cite{Altendorf1993}, further corroborating the significance of EPC in this system.

Finally, we discuss the interplay between the CDW order and magnetism. The dispersion dips at $\boldsymbol{q}_\mathrm{I}$ and $\boldsymbol{q}_\mathrm{C}$ return/persist into the ferromagnetic phase, where they are even more pronounced than at ambient conditions (Fig. \ref{Fig4}). This reveals that ferromagnetism destroys the CDW order but not the underlying tendency towards ordering, despite the reduction of Fermi-surface nesting and susceptibility peaks~\cite{Laverock2009,Kim2013}. The EPC adjusts to the spin-polarization of the electronic bands, as indicated also by the phonon resharpening at $\boldsymbol{q}_\mathrm{C}$ below $T_\mathrm{C}$. However, our observations indicate that locally it remains sufficiently strong to preserve both ordering tendencies, yet not strong enough to stabilize one of the CDWs. While SmNiC$_2$ is unique in its extreme interplay of charge and magnetic order, antiferromagnetism in other $R$NiC$_2$ members also affects their CDW states ~\cite{Yamamoto2013,Shimomura2016,Maeda2019}. Although antiferromagnetic order can also influence the nesting properties, our results highlight that changes in the EPC in the magnetic states should be considered an important contributor to the phase interplay.

\textit{Conclusions --} Our data establish that the I-CDW transition in SmNiC$_2$ is soft-phonon driven and that a Kohn anomaly also appears around the wavevector of the C-CDW stabilized in other $R$NiC$_2$ members. We observe a competitive trend between the two CDW ordering tendencies, as well as the persistence of both in the ferromagnetic state, despite the suppression of long-range order and the associated Fermi surface nesting features. The observed collective AM of the CDW directly mirrors the formation and suppression of the I-CDW and -- together with the pronounced phonon renormalizations -- hints at the presence of significant EPC. Our observations underscore that a synergy between Fermi surface topology and EPC governs the CDW stabilization in this system. More generally, our results highlight the significance of EPC and its evolution across electronic transitions in shaping the landscape of competing phases.

\textit{Acknowledgments --} We acknowledge the European Synchrotron Radiation Facility (ESRF) for provision of synchrotron radiation facilities under proposal numbers HC-4325 and HC-5552. We thank F. Henßler, T. Lacmann, M.-A. Méasson, F. Weber and M. Ye for fruitful discussions. This work was funded by the Deutsche Forschungsgemeinschaft (DFG, German Research Foundation) – Projektnummer 441231589. R.~H. acknowledges support by the state of Baden-W\"{u}rttemberg through bwHPC.

\bibliography{SmNiC2}

\end{document}